\documentstyle[12pt]{article}
\input{tcilatex}

\begin{document}

\title{{\huge Interface disorder and layer transitions in Ising thin films}}
\author{{\bf L. Bahmad, and A. Benyoussef} \\
%EndAName
Laboratoire de Magn\'{e}tisme et de la Physique des Hautes Energies \\
Universit\'{e} Mohammed V, Facult\'{e} des Sciences, Avenue Ibn Batouta, \\
Rabat B.P. 1014, Morocco}
\date{}
\maketitle

\begin{abstract}
\mbox{~~~} The disorder and layer transitions in the interface between an
Ising spin-1/2 film denoted $(n)$, and an Ising spin-1 film denoted $(m)$,
are studied using Monte Carlo simulations. The effects of both an external
magnetic field, acting only on the spin-1/2 film , and a crystal magnetic
field acting only on the spin-1 film, are studied for a fixed temperature
and selected values of the coupling constant $J_p$ between the two films. It
is found that for large values of the constant $J_p$, the layers of the film 
$(n)$, as well as those of the film $(m)$, undergo a first order layering
transition. On the other hand, the only disordered layer of the film $(n)$
is that one belonging to the interface films $(n)$/$(m)$, for any values of
the crystal field $\Delta$. We show the existence of a critical value of the
crystal field $\Delta_c$, above which this particular layer of the film $(n)$
is disordered. We found that $\Delta_c$ depends on the values of the
constant coupling $J_p$ between the two films. \newline
\end{abstract}

----------------------------------- \newline
\mbox{~}(*) Corresponding author: bahmad@fsr.ac.ma \newline

%\newpage

\section{Introduction}

\mbox{  } Layered structures consisting of alternating magnetic materials,
have been subject of many experimental study. These experimental studies
have shown that the magnetisation enhancement exists in multilayered films
consisting of magnetic layers. The most commonly studied magnetic
multilayers are those of ferromagnetic transition metal such as $Co$ or $Ni$%
. Several experiments have shown that the magnetisation enhancement exists
in multilayered films consisting of magnetic layers. It was found that
ferromagnetic coupling can exist between magnetic layers. From the
theoretical point of view, great interest has been paid to spin wave
excitations as well as critical phenomena [1-3]. The study of thin films is
partly motivated by the development of new growth and characterisation
techniques, but perhaps more so by the discovery of many exciting new
properties, some quite unanticipated. These include, more recently, the
discovery of enormous values of magnetoresistance in magnetic multilayers
far exceeding those found in single layer films and the discovery of
oscillatory interlayer coupling in transition metal multilayers. \newline
These experimental studies have motivated much theoretical work. Indeed,
several methods have been used to study the layering transitions in Ising
magnetic films. Benyoussef and Ez-Zahraouy have studied these layering
transitions using a real space renormalization group [4], and transfer
matrix methods [5]. Using the mean field theory, Hong [6], have found that
depending on the values of the exchange integrals near the surface region,
the film critical temperature may be lower, higher than, or equal to that of
the bulk. Using the perturbative theory, Harris [7] have showed the
existence of layering transitions at $T=0$ in the presence of a transverse
magnetic field. The effect of finite size on such transitions has been
studied, in a thin film confined between parallel plates or walls by Bruno ${\it et al.}$
[8] taking into account the capillary condensation effect. By applying
Monte Carlo simulations on thin Ising films with competing walls, Binder 
{\it et al.} [9], found that occurring phase transitions belong to the
universality class of the two-dimensional Ising model and found that the
transition is shifted to a temperature just below the wetting transition of
a semi-infinite fluid [10,11]. Hanke ${\it et al.}$ [12] showed that
symmetry breaking fields give rise to nontrivial and long-ranged order
parameter profiles for critical systems such as fluids, alloys, or magnets
confined to wedges. \newline
We showed in one of our earlier works [13] the existence of layering
transitions under the effect of a variable surface coupling. Moreover, for
an Ising film with a wedge, we found in Ref. [14] the intra-layering
transitions under the geometry effect consisting on the existence a wedge.
When the film is subject to a random transverse magnetic field [15], we
found the layer-by-layer transitions when increasing the concentration $p$
above a critical value $p_c(k)$ for each layer $k$. The aim of this work is
to study the interface disorder coupled to layering transitions in a spin$%
-1/2$ film in interaction with a spin$-1$ film, using Monte Carlo
simulations. The paper is organised as follows. In section $2$, we give the
model and the method used. In section $3$ we present results and discussions.

\section{Model and Monte Carlo simulations}

We are studying two coupled magnetic films: a spin-1/2 film denoted $(n)$
formed with $n$ square ferromagnetic layers; and a spin-1 film denoted $(m)$
formed with $m$ square ferromagnetic layers, Fig. 1. Each layer is a square
of dimension $N_x \times N_y=64 \times 64$ spins. $N_x$ and $N_y$ being the
number of spins in the $x$ and $y$ directions, respectively. The juxtaposed
plans of the two films are coupled ferromagnetically via a constant $J_p$,
so that the Hamiltonian governing the system is given by: 
\begin{equation}
{\cal H}=-J\sum_{<i,j>_{(n)}}S_{i}S_{j}-H \sum_{i \epsilon
(n)}S_{i}-J\sum_{<i,j>_{(m)}}\sigma_{i}\sigma_{j}-J_p\sum_{i \epsilon (n), j
\epsilon (m)}S_{i}\sigma_{j}+\Delta \sum_{j \epsilon (m)}\sigma_{j}^2
\end{equation}
where, $S_{l}(l=i,j)=-1,+1$ are the spin variables in the film $(n)$ and $%
\sigma_{k}(k=i,j)=0,\pm 1$ are the spin variables in the film $(m)$. The
spins of the film $(n)$ are subject to an external magnetic field $H$,
whereas $\Delta$ is the crystal field acting only on spins of the film $(m)$%
. The sum $\sum_{<i,j>_{(n)}}$ (resp. $\sum _{<i,j>_{(m)}}$) is performed on
nearest neighbour spins of the film $(n)$ (resp. of the film $(m)$). The
interaction $J$ between the spins of the film $(n)$ as well as between the
spins of the film $(m)$, is assumed to be constant. Values of the interface
coupling constant $J_p$ will be discussed in all the following. A
preliminary study showed that, when performing Monte Carlo simulations under
the Metropolis algorithm,  the relevant calculated quantities did not change
appreciably for the film thickness: $n,m=3,4,5$ and $8$ layers; and when
varying the number of spins of each layer from $N_x=N_y=32$ to $128$. Where $%
N_x$ and $N_y$ are the number of spins, of each layer, in the $x$ and $y-$%
directions, respectively. Taking into account the above considerations, in
all the following we will give numerical results for a film thickness $n=m=5$
layers, and $N_x=N_y=64$ spins for each layer. \newline

\section{Results and discussion}

\mbox{~~~} The geometry of the system we are studying is summarised in Fig.
1, where the two films are coupled via the constant $J_p$. Each layer of the
film $(n)$ is subject to the external magnetic field $H$; whereas the layers
of the film $(m)$ are under the crystal field $\Delta$. Although the model
and established equations are valid for arbitrary values number of layers $n$
and $m$, numerical results will be given in this work for $n=m=5$ layers.
Also the temperature fluctuations will not be taken into account and will be
fixed at the value $T=3.0$ in all the following. \newline
In order to outline the effect of the external magnetic field, we plot in
Figs. $2a$ and $2b$ the layer magnetisation behaviour in each film: $n$ and $%
m$ for a fixed crystal field $\Delta=2.0$. The former figure, plotted for a
small coupling constant value $J_p=0.5$, shows that only the layers of film $%
(n)$ (spin$-1/2$) transit when increasing the external magnetic field,
including the interface layer $n(1)$. Hence for small values of the
parameter $J_p$ the layering transitions are absent in the film $(m)$ (spin$%
-1$) for all values of the external magnetic field. But for higher values of
the coupling constant $J_p$ the layering transitions begin to occur in the
film $(m)$, as it is shown in Fig. $2b$ for $J_p=9.0$. When increasing the
amplitude of $J_p$ more and more all the layers of the film $(m)$ transit,
following the layering transitions present in the film $(n)$. This means
that for sufficiently large values of the coupling constant $J_p$, the
effect of the magnetic field $H$ acting on the film $(n)$ is propagated to
the film $(m)$ layers. \newline
On the other hand, one can ask if the crystal field $\Delta$, applied on the
film $(m)$ and disordering its layers, can affect the layers of the film $n$%
. To answer this question, we plot in Figs. $3a$ and $3b$ the layer
magnetisation behaviour in each film as a function of the crystal field $%
\Delta$ for a fixed value of the external magnetic field $H=0.20$ and two
values of the coupling constant $J_p$. Indeed, the only disordering layer of
the film $(n)$ is that one belonging to the interface: $n(5)$. The disorder
of this specific layer is affected neither by small values of the coupling
constant: $J_p=0.5$ in Fig. $3a$; nor by higher values: $J_p=9.0$ in Fig. $3b
$. Hence, the disorder of the film $(m)$ does not affect the layers of the
film $(n)$ except the special layer directly connected to the interface
between the two films. \newline
On the other hand, the effect of the interface coupling constant is
summarised in Figs. $4a$ and $4b$ corresponding to a low crystal field value 
$\Delta=0.5$ and a higher value $\Delta=9.0$, respectively. It is found that
starting from a totally disordered film $(m)$ for low values of $\Delta$,
when increasing the coupling constant $J_p$ values, all the layers of the
film $(m)$ undergo non null magnetisations (keeping the sign of the film $(n)
$ layers), see Fig. $4a$. While for higher values of the crystal field $%
\Delta$, see Fig. $4b$ , the increasing coupling $J_p$ values does not
affect, neither disorder of the film $(m)$ layers, nor the layer
magnetisations of the film $(n)$. Indeed, the special layer $n(5)$
disordered for the high crystal field value $\Delta=9.0$ at low values of
the coupling $J_p$, is not affected by increasing the coupling $J_p$ between
the two films. \newline
In the following we will focus our interest on the particular layers of the
interface of the films: $n(5)$ belonging to the spin-$1/2$ film $(n)$, and $%
m(1)$ belonging to the spin-$1$ film $(m)$. Indeed, concerning the
transition of the layers $n(5)$ for a fixed crystal field value $\Delta=1.0$
we show in Fig. $5a$ the effect of increasing coupling $J_p$ on this layer.
It is shown that the transition of the layer $n(5)$ inside the film $(n)$ is
not affected by increasing the coupling $J_p$ from low values $0.1$, $1.0$
to a higher value $9.0$. Whereas, under the same conditions, the transition
of the layer $m(1)$ is strongly affected by increasing the coupling $J_p$,
See Fig. $5b$. This layer does not transit for $J_p=0.1$ and any value of
the external magnetic field $H$. \newline
On the other hand, the layer $n(5)$ disorders for $\Delta \ge 9.0$ , see
Fig. $6a$, for any coupling $J_p$ value; whereas the layer $m(1)$ disorders
for $J_p=0.1$ at $\Delta \approx 2.0$, for $J_p=1.0$ at $\Delta \approx 6.0$
and for $J_p=9.0$ at $\Delta \ge 10.0$, Fig. $6b$. This means that the
crystal field needed to disorder these layers depends on the coupling $J_p$
of the interface between the two films. A scenario of this dependence is
plotted in Fig. $7$ for the layer $n(5)$ and a fixed external magnetic field 
$H=0.20$. It is seen, from this figure, that to disorder the layer $n(5)$
for low values of the coupling $J_p$, small values of the crystal field are
needed. When increasing the coupling $J_p$, the crystal field $\Delta_c$
undergoes a maximum value at $\approx 9.0$, and decreases when increasing
the coupling $J_p$ more and more.

\section{Conclusion}

The disorder and transitions of The interface between an Ising spin-1/2 film
denoted $(n)$, and an Ising spin-1 film denoted $(m)$, has been studied
using Monte Carlo simulations. The effects of both an external magnetic
field, responsible on the interface transition, and a crystal magnetic field
needed to disorder the interface are studied for a fixed temperature. It is
found that for low values of the coupling $J_p$, the layers of the film $(n)$
undergo a first order layering transition. These transitions are also found
in the film $(m)$, for higher values of the coupling $J_p$ between the two
films. On the other hand, the layers of the film $(m)$ disorder when
increasing the crystal magnetic field for any coupling $J_p$ value, but the
only disordering layer of the film $(n)$ is that one belonging to the
interface between the films $(n)$/$(m)$. \newline
On the other hand, we show the existence of a critical value of the crystal
field $\Delta_c$, above which this particular layer of the film $(n)$ is
disordered at given values of the interface coupling $J_p$. \newline
\noindent {\Large {\bf References}}

\begin{enumerate}
\item[\lbrack 1]]  S. Dietrich and M. Schick, Phys. Rev B {\bf 31},4718
(1985)

%\item[\lbrack 2]]  S. J. Kennedy and S. J. Walker, Phys. Rev. B {\bf 30}%
%,1498 (1984)

\item[\lbrack 2]]  P. Wagner and K. Binder, Surf. Sci. {\bf 175},421 (1986)

\item[\lbrack 3]]  K. Binder and D. P. Landau, Phys. Rev. B {\bf 37}, 1745
(1988)

\item[\lbrack 4]]  A. Benyoussef and H. Ez-Zahraouy, Physica A, {\bf 206},
196 (1994).

\item[\lbrack 5]]  A. Benyoussef and H. Ez-Zahraouy, J. Phys. {\it I }
France {\bf 4}, 393 (1994).

\item[\lbrack 6]]  Q. Hong Phys. Rev. B {\bf 41}, 9621 (1990); ibid, Phys.
Rev. B {\bf 46}, 3207 (1992).

\item[\lbrack 7]]  A. B. Harris, C. Micheletti and J. Yeomans, J. Stat.
Phys. {\bf 84}, 323 (1996)

%\item[\lbrack 9]]  H. Nakanishi and M. E. Fisher, J. Chem. Phys. {\bf 78}%
%,3279 (1983)

\item[\lbrack 8]]  P. S. Swain and A. O. Parry, Eur. Phys. J. {\bf B 4},
459 (1998); E. Bruno, U. Marini, B. Marconi and R. Evans, Physica A {\bf 141A%
}, 187 (1987)

\item[\lbrack 9]]  K. Binder, D. P. Landau and A. M. Ferrenberg, Phys. Rev.
Lett. {\bf 74}, 298 (1995)

\item[\lbrack 10]]  K. Binder, D. P. Landau and A. M. Ferrenberg, Phys. Rev. 
{\bf E 51}, 2823 (1995)

\item[\lbrack 11]]  M. Bengrine, A. Benyoussef, H. Ez-Zahraouy and F.
Mhirech, Physica {\bf A 268}, 149 (1999)

\item[\lbrack 12]]  A. Hanke, M. Krech, F. Schlesener and S. Dietrich, Phys.
Rev. E {\bf 60}, 5163 (1999)

\item[\lbrack 13]]  L. Bahmad, A. Benyoussef, and H. Ez-Zahraouy, Surf. Sci. 
{\bf 536}, 114 (2003)

\item[\lbrack 14]]  L. Bahmad, A. Benyoussef, and H. Ez-Zahraouy, Phys. Rev.
E {\bf 66}, 056117 (2002)

\item[\lbrack 15]]  L. Bahmad, A. Benyoussef, and H. Ez-Zahraouy, Chin. J.
Phys. {\bf 40}, 537 (2002)
\end{enumerate}

%\newpage
\noindent{\bf Figure Captions}\newline

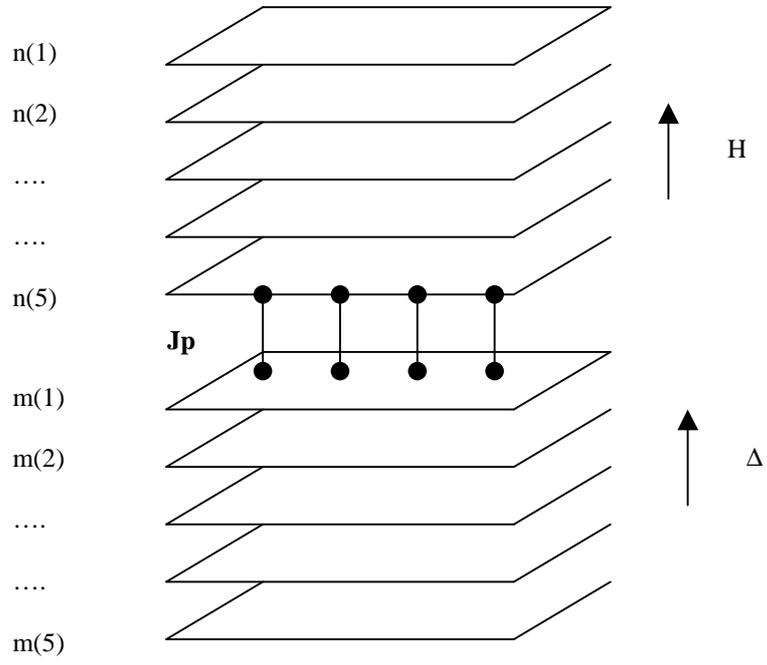
\noindent{\bf Figure 1.}: \newline

Geometry of the system formed with two films coupled via the constant $J_p$.
Each layer of the film $(n)$ is subject the external magnetic field $H$;
whereas the crystal field $\Delta$ is acting only on the layers of the film $%
(m)$. The films $(n)$ and $(m)$ are formed with the same number of layers, $%
N=5$. \newline

\noindent{\bf Figure 2}: \newline
The external magnetic field effect on the layer magnetisation in each film,
for a fixed crystal field $\Delta=2.0$. (a) For a small coupling constant
value $J_p=0.5$, only the layers of film $(n)$ transit when increasing the
external magnetic field, including the interface layer $n(5)$. The layering
transitions are absent in the film $(m)$ for this small value of the
coupling constant. (b) For a higher value of the coupling constant $J_p=9.0$
the layering transitions begin to occur in the film $(m)$ following the
layering transitions present in the film $(n)$. \newline

\noindent{\bf Figure 3}: \newline
The layer magnetisation behaviour in each film as a function of the crystal
field $\Delta$ for $H=0.20$. (a) For a small value of the coupling constant: 
$J_p=0.5$, when increasing $\Delta$ all the layers of the film $(m)$ are
disordered but the only disordering layer of the film $(n)$ is that one
belonging to the interface: $n(5)$. (b) For a higher value of the coupling
constant: $J_p=9.0$, the disorder of the film $(m)$ layers does not affect
the layers of the film $(n)$ except the special layer $n(5)$ directly
connected to the interface between the two films. \newline

\noindent{\bf Figure 4}:\newline
The interface coupling effect on the layer magnetisation behaviour in each
film for a constant external magnetic field $H=0.20$ and two crystal field
values: (a) For $\Delta=0.5$ and starting from a totally disordered film $(m)
$ all the layers of the film $(m)$ undergo non null magnetisations when the
coupling $J_p$ between the films is increasing. (b) For a higher crystal
field value $\Delta=9.0$, the increasing coupling $J_p$ values does not
affect, neither the disorder of the film $(m)$ layers, nor the layer
magnetisations of the film $(n)$, except the special layer $n(5)$ belonging
to the interface. \newline

\noindent{\bf Figure 5}: \newline
Magnetisation behaviour of the interface layers: $n(5)$ and $m(1)$ for a
fixed crystal field value $\Delta=1.0$ and different values of the coupling
constant $J_p=0.1$, $1.0$ and $9.0$, as a function of the external magnetic
field $H$. (a) The transition of the layer $n(5)$ inside the film $(n)$ is
not affected by increasing the coupling constant. (b) The layer $m(1)$ does
not transit for low values of the coupling, e.g. $J_p=0.1$. The transition
of this layer is only seen for large values of the coupling $J_p$. \newline

\noindent{\bf Figure 6}: \newline
Magnetisation behaviour of the interface layers: $n(5)$ and $m(1)$ for a
fixed external magnetic field value $H=0.20$ and different values of the
coupling constant $J_p=0.1$, $1.0$ and $9.0$, as a function of the crystal
field $\Delta$. The crystal field needed to disorder the layer $n(5)$ (a)
and the layer $m(1)$ (b) depends strongly on the coupling $J_p$ between the
two films. \newline

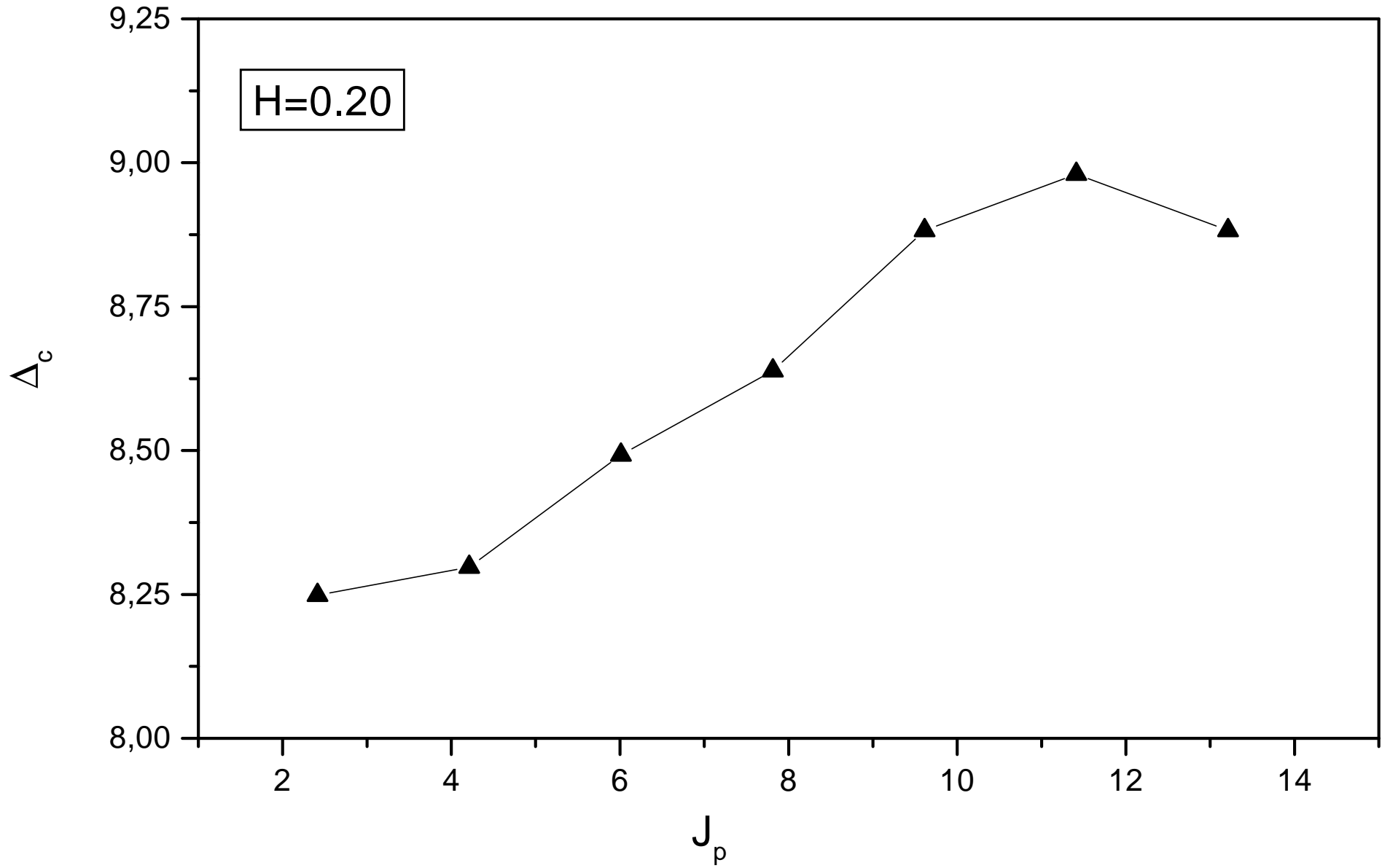
\noindent{\bf Figure 7}: \newline
The critical crystal field $\Delta_c$ needed to disorder the layer $n(5)$ as
a function of the coupling $J_p$, for a fixed external magnetic field $H=0.20
$. $\Delta_c$ presents a maximum value for a coupling $J_p \approx 12.0$ and
decreases when increasing the coupling $J_p$ between the two films.

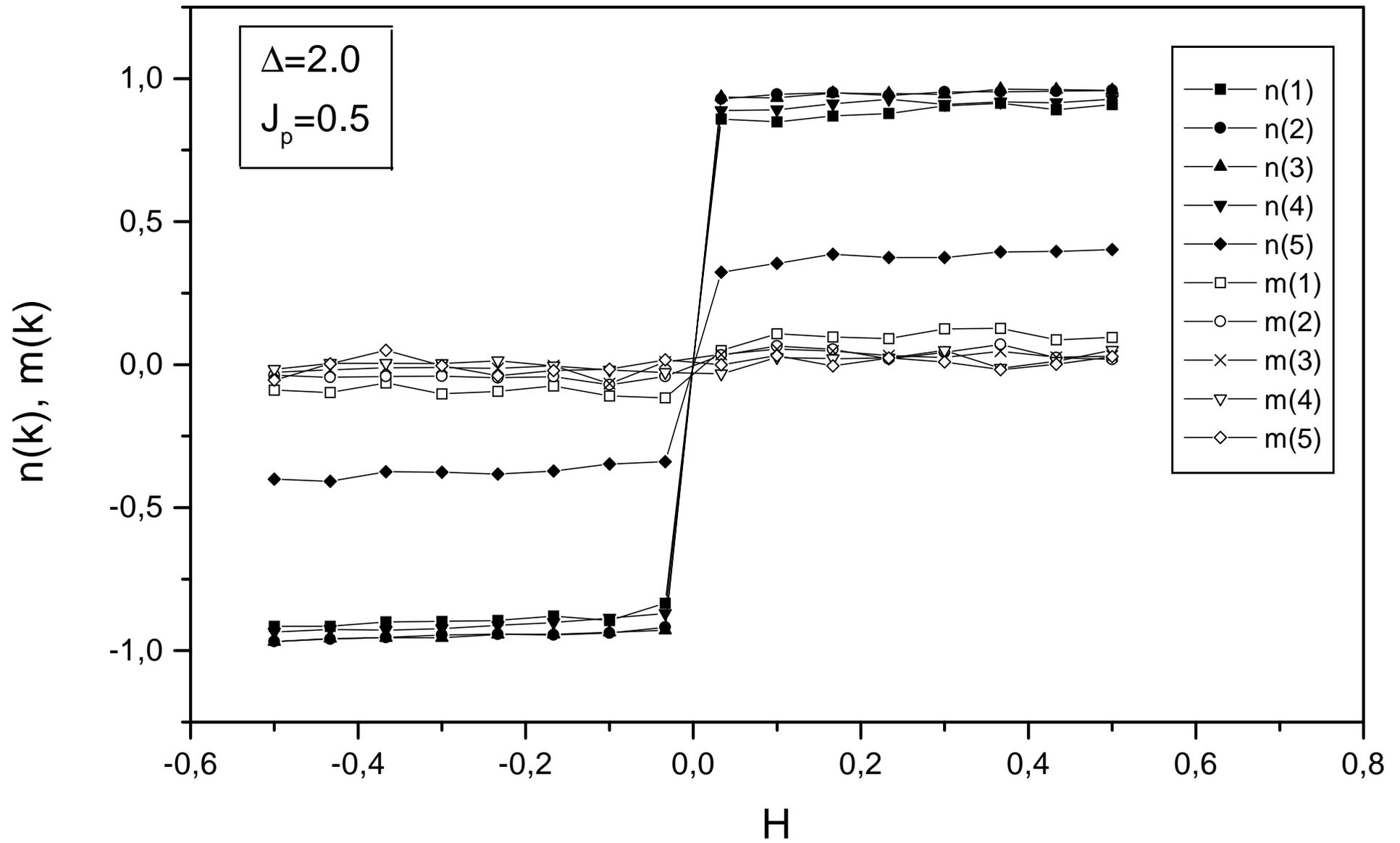
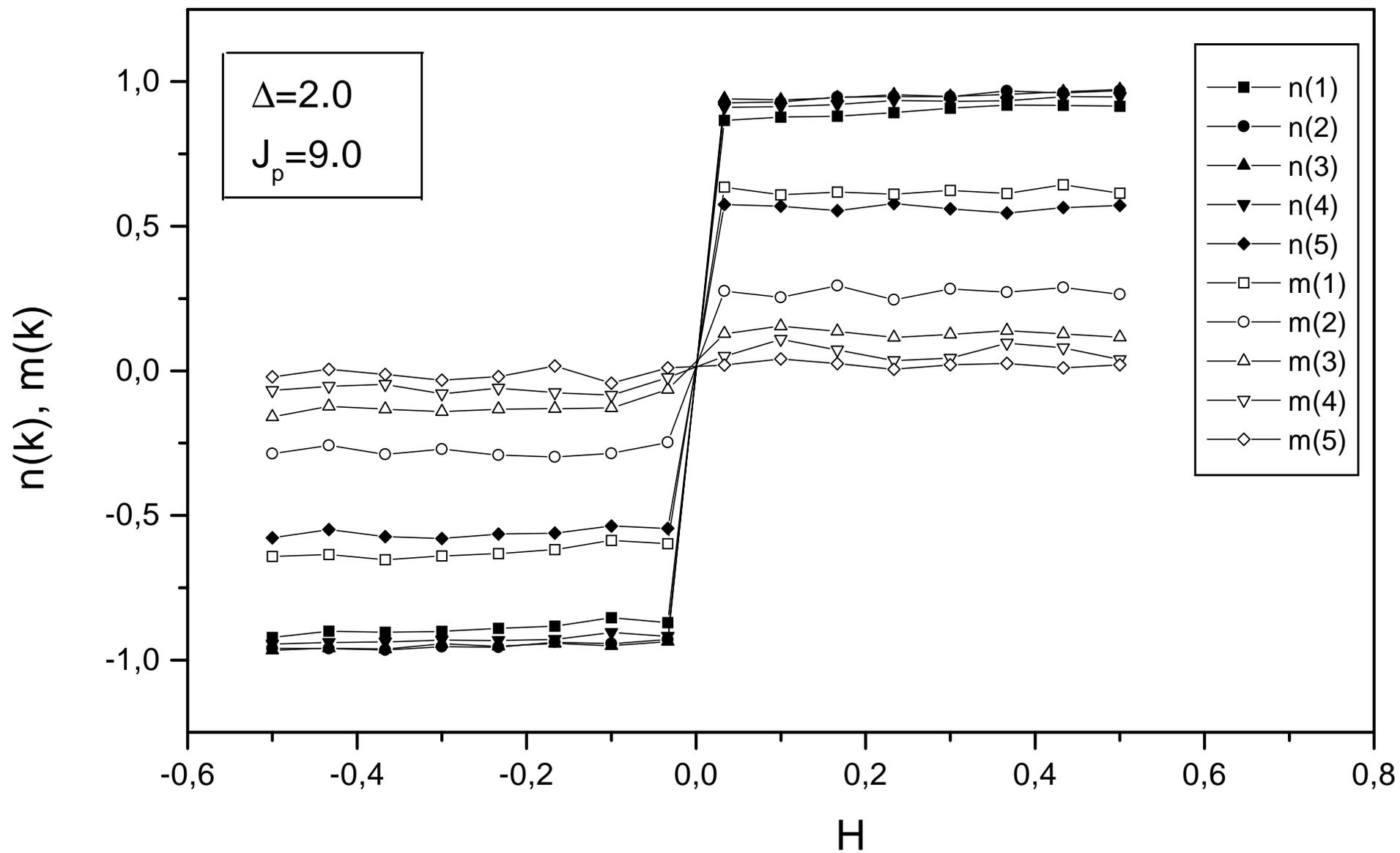
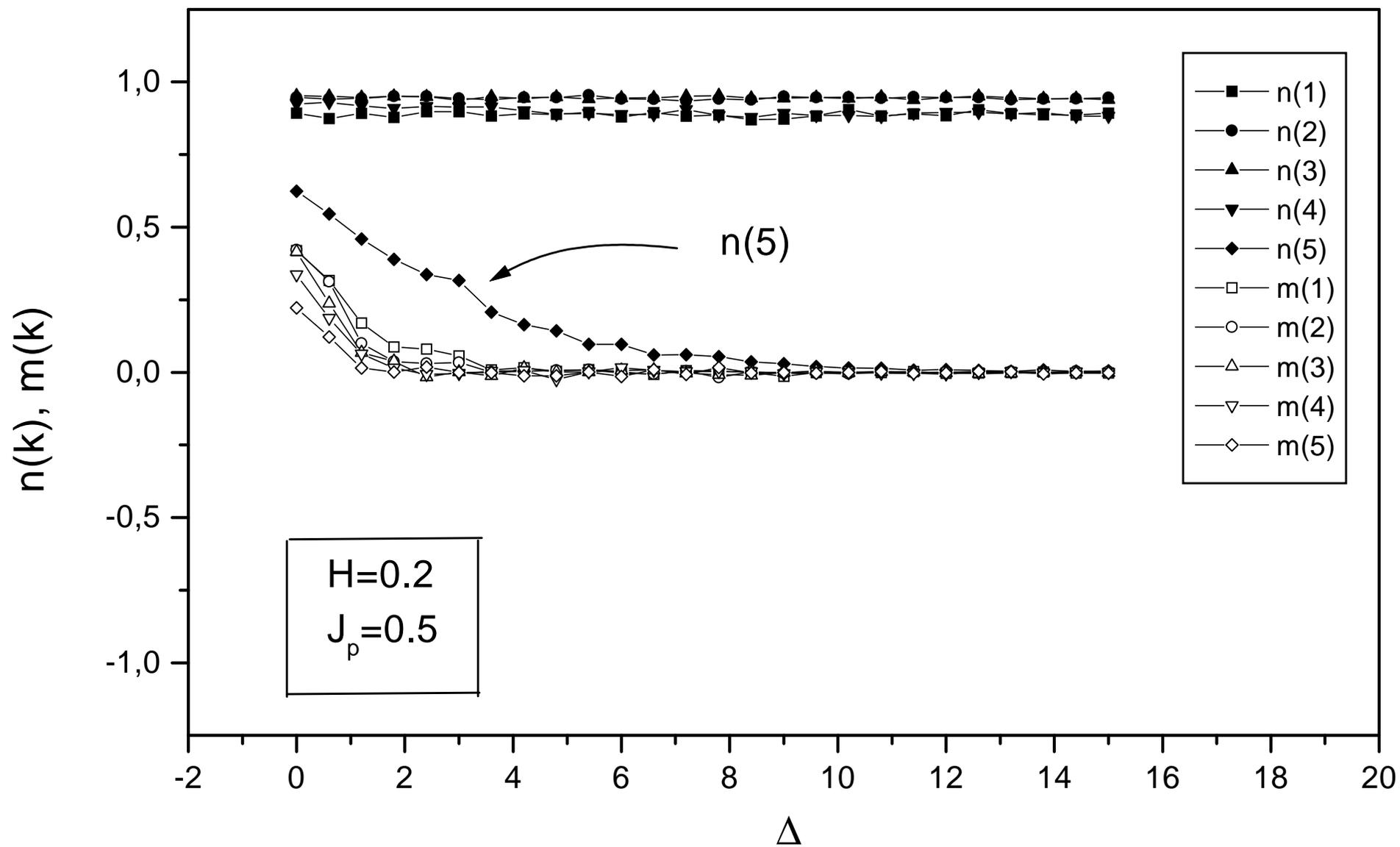
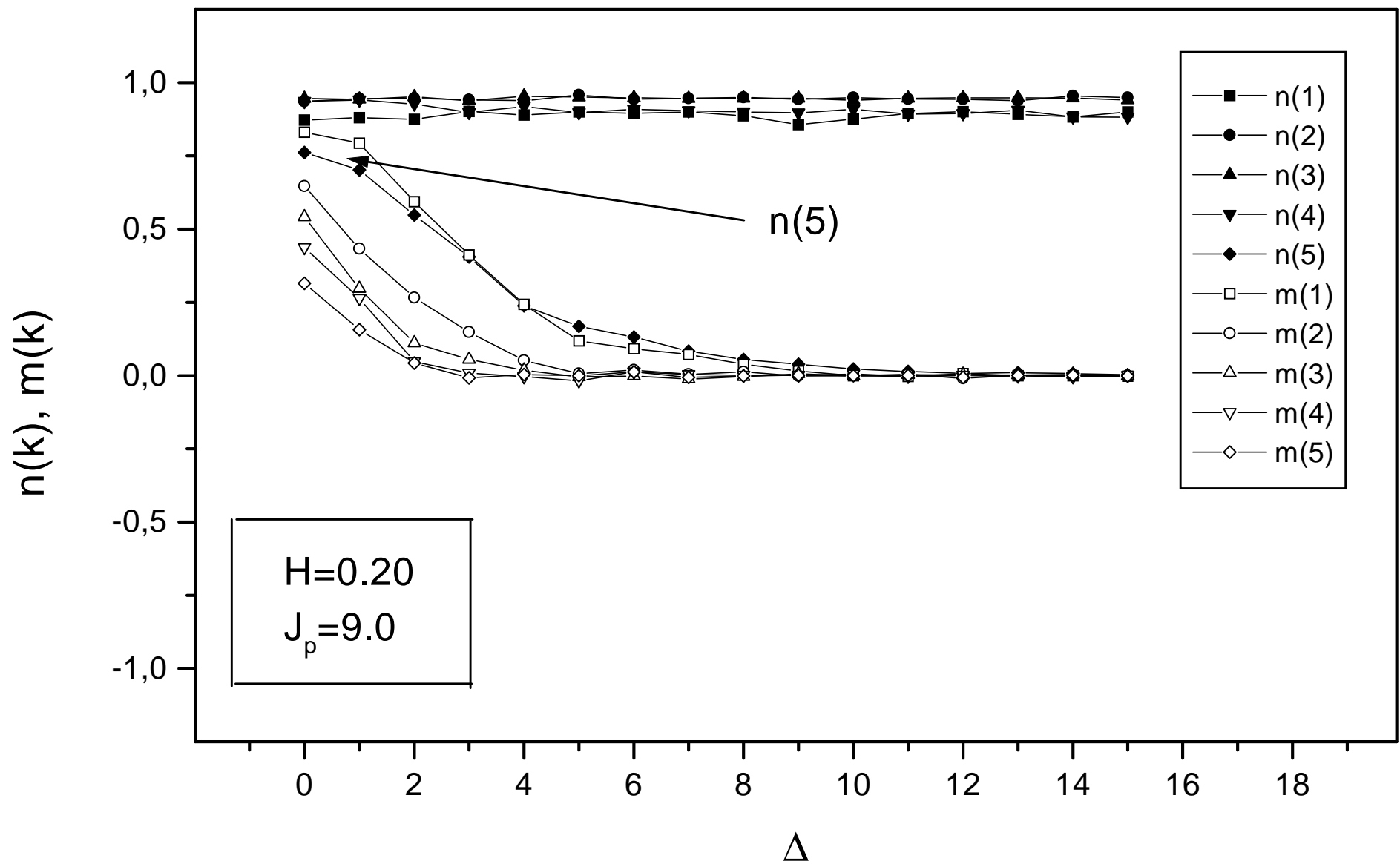
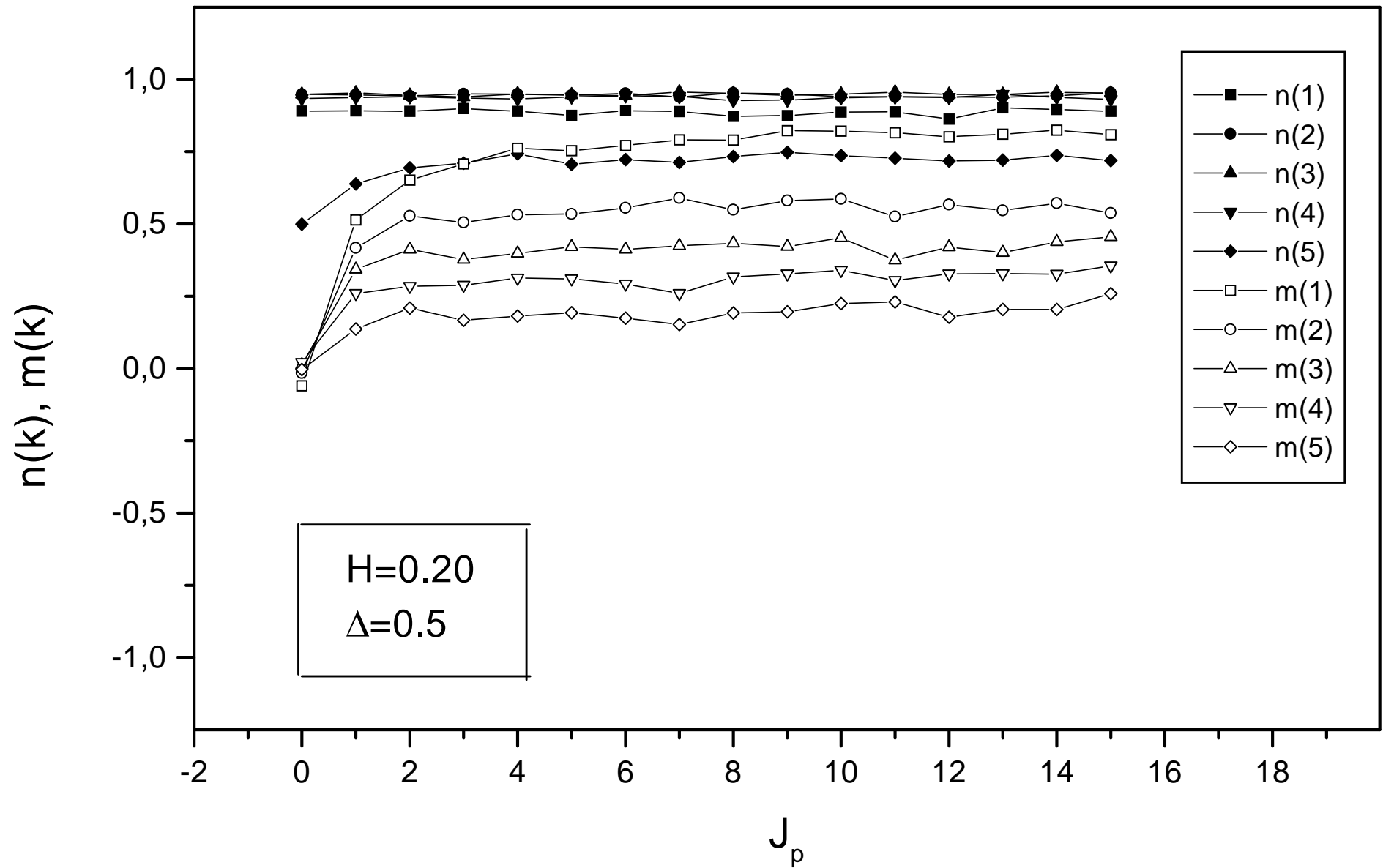
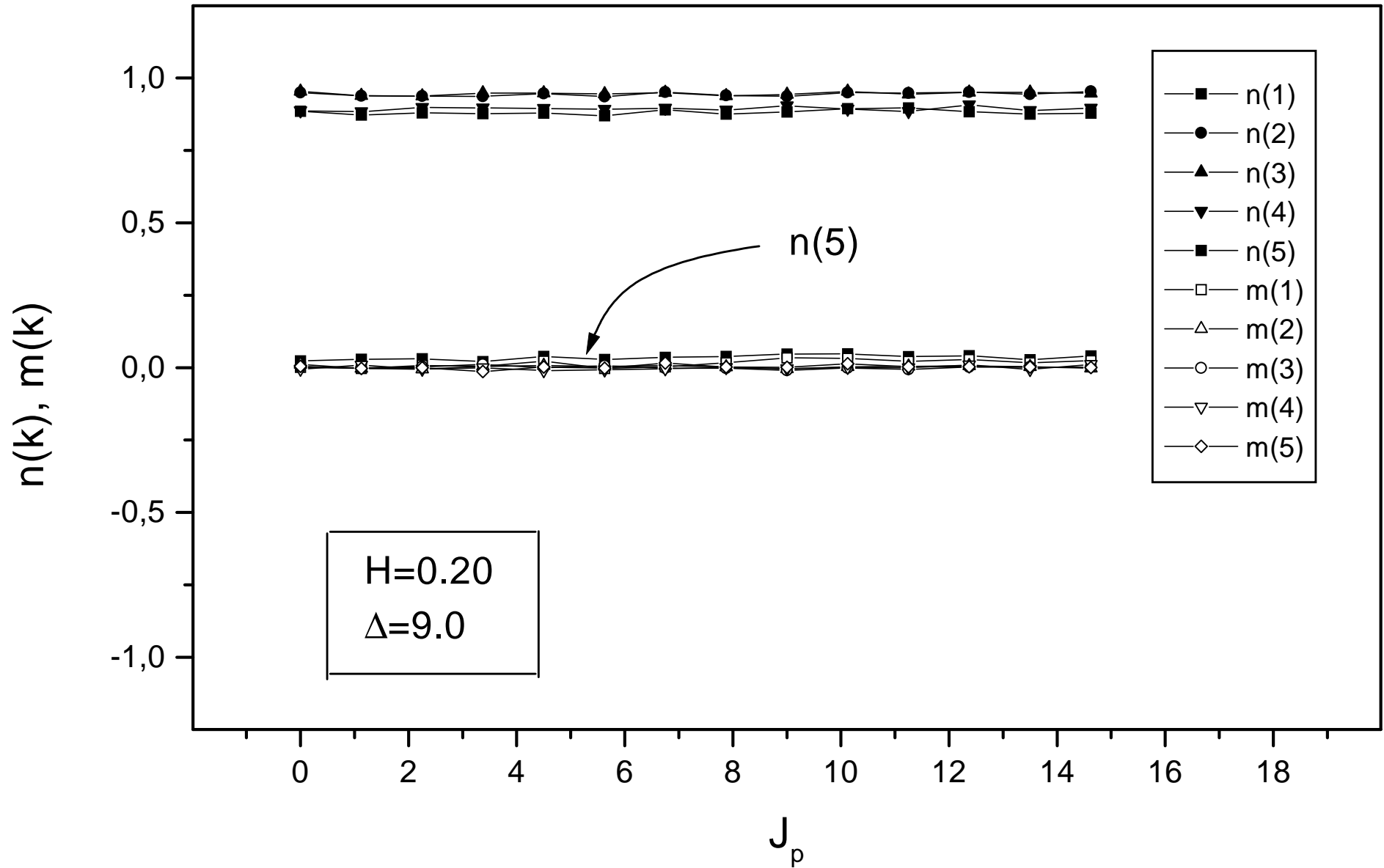
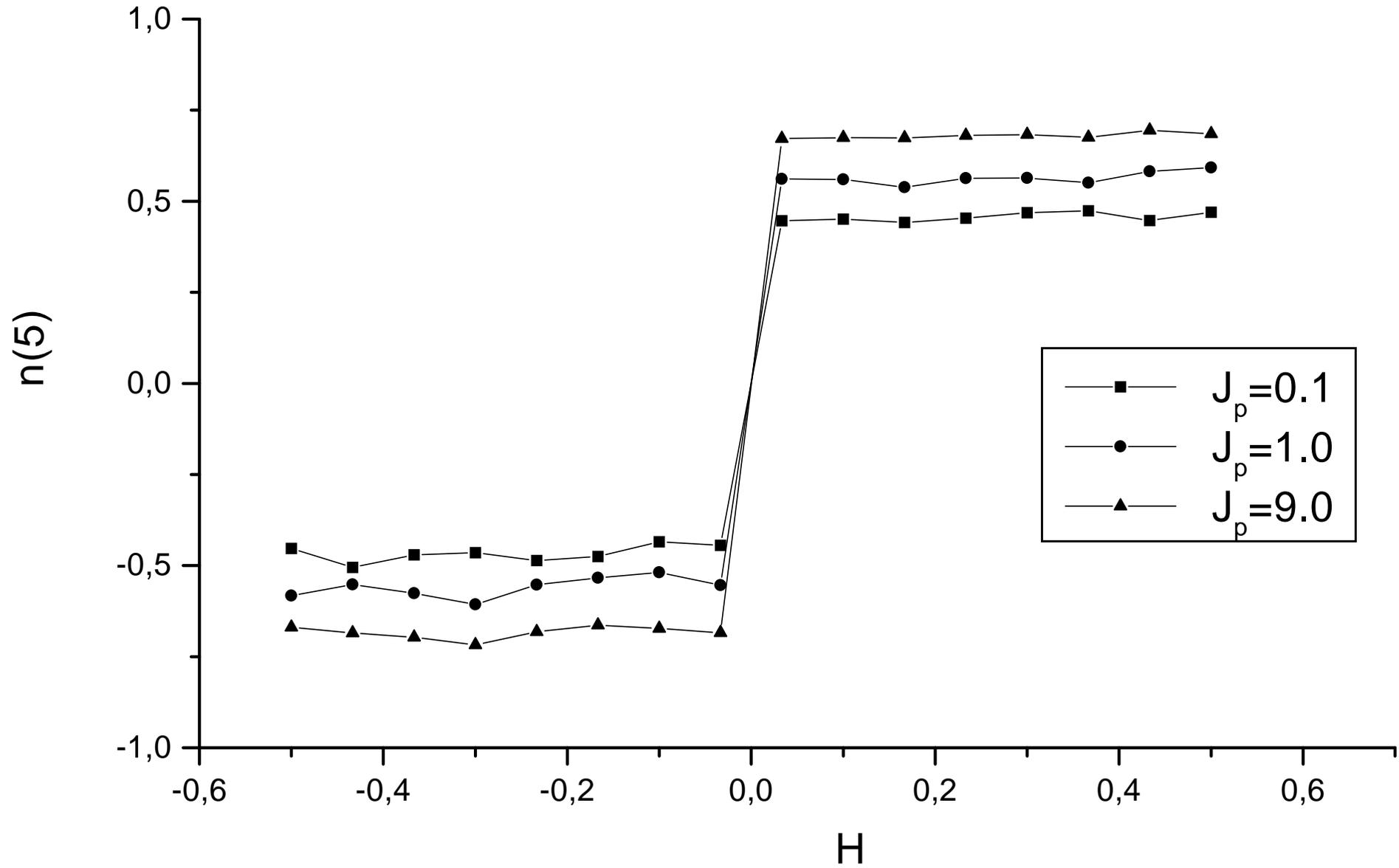
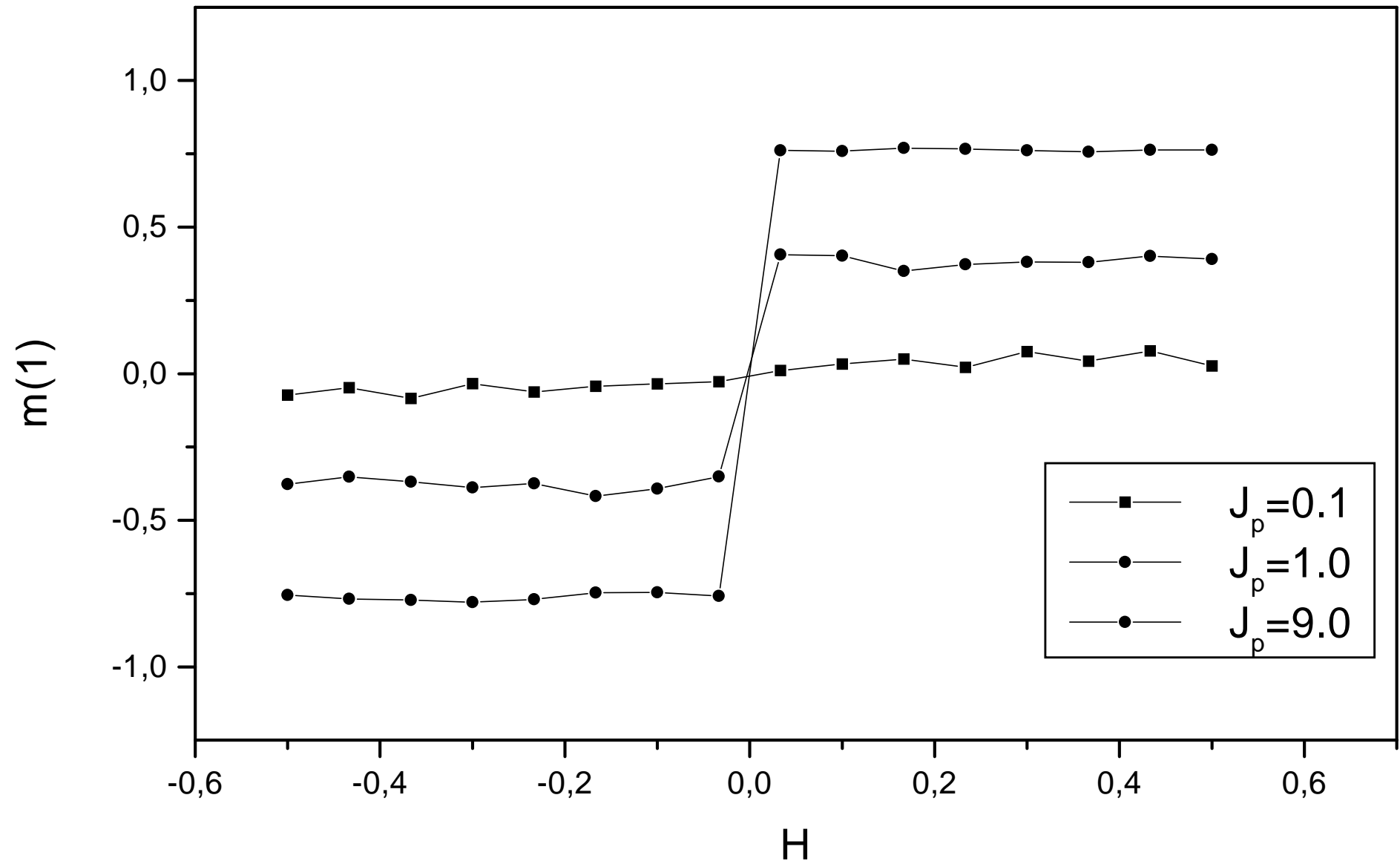
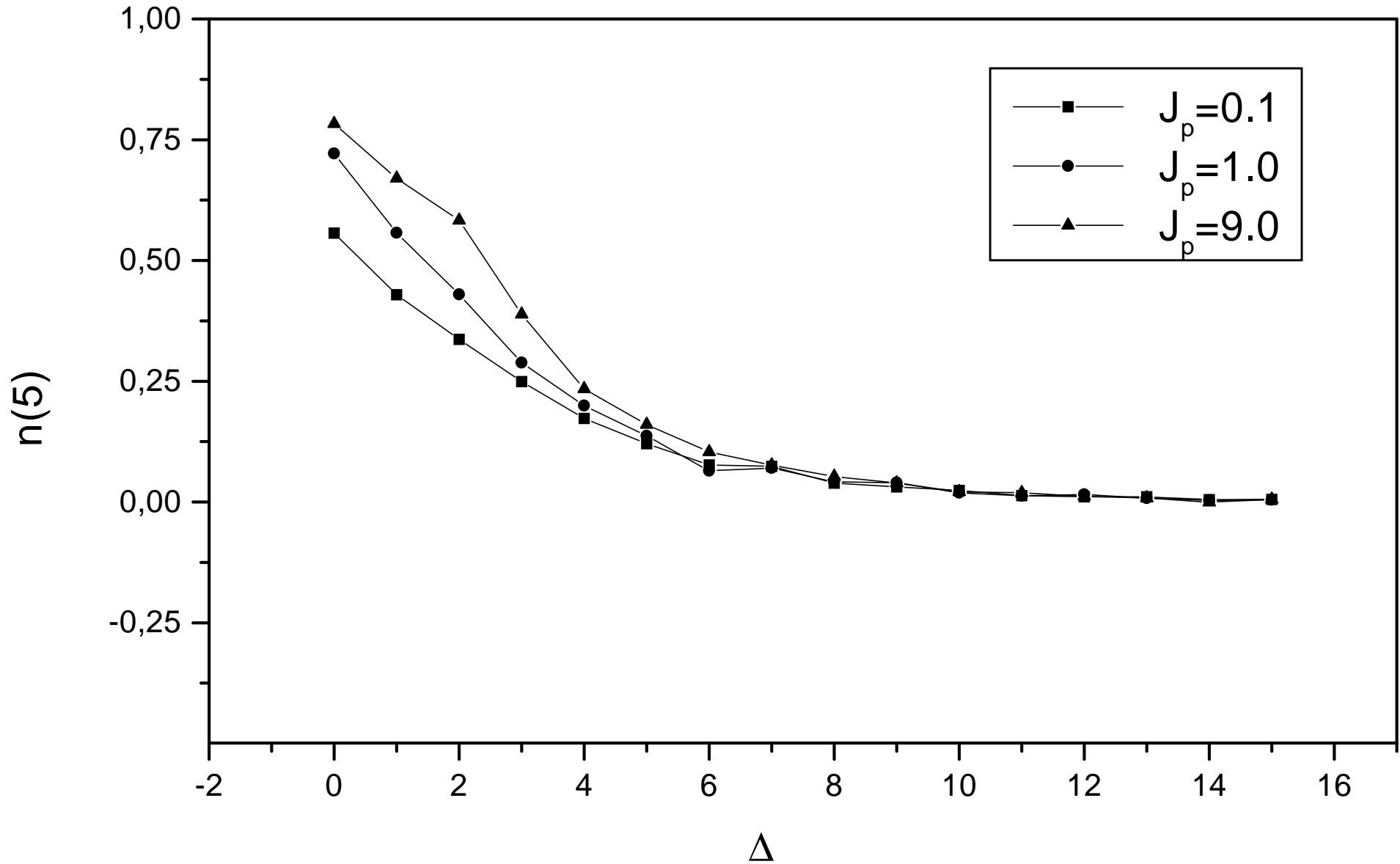
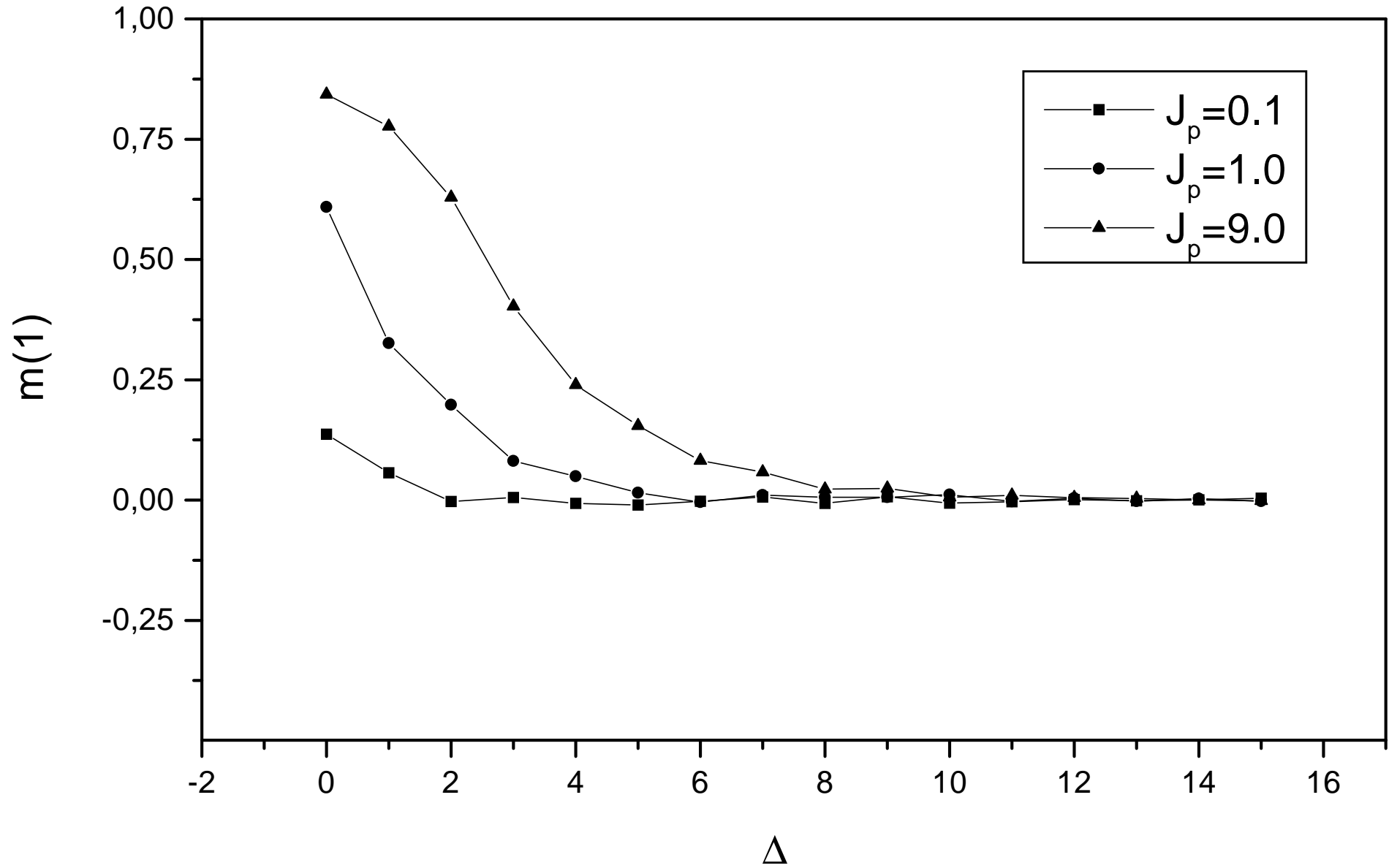
\end{document}